\documentclass{aastex63}

\received{2021 February 27}
\revised{2021 April 12}
\accepted{2021 May 2}

\submitjournal{ApJ}

\shorttitle{Interaction of a Bonnor-Ebert sphere with a stellar wind}
\shortauthors{Zier et. al}
\usepackage{color}
\usepackage{amssymb}
\usepackage{pifont}
\usepackage{amsmath}
\usepackage{mathtools}
\usepackage{lineno}
\begin{document}

\title{On the interaction of a Bonnor-Ebert sphere with a stellar wind}

\correspondingauthor{Oliver Zier}
\email{ozier@mpa-garching.mpg.de}

\author[0000-0003-1811-8915]{Oliver Zier}
\affiliation{Max-Planck-Institut f\"ur Astrophysik (MPA), 
Karl-Schwarzschild-Strasse 1,
85748 Garching, Germany}

\author[0000-0001-6879-9822]{Andreas Burkert}
\affiliation{University Observatory Munich (USM),
Scheinerstrasse 1,
81679 Munich,Germany}
\affiliation{Max-Planck-Institut f\"ur extraterrestrische Physik (MPE),
Giessenbachstr. 1,
85748 Garching, Germany
}

\author{Christian Alig}
\affiliation{University Observatory Munich (USM),
Scheinerstrasse 1,
81679 Munich,Germany}

\DeclarePairedDelimiter\ceil{\lceil}{\rceil}
\DeclarePairedDelimiter\floor{\lfloor}{\rfloor}
\newcommand\msun{\rm M_{\odot}}
\newcommand\lsun{\rm L_{\odot}}
\newcommand\msunyr{\rm M_{\odot}\,yr^{-1}}
\newcommand\msunMyr{\rm M_{\odot}\,Myr^{-1}}
\newcommand\msunpc{\rm M_{\odot}\,pc^{-3}}
\newcommand\Myr{\rm Myr}
\newcommand\be{\begin{equation}}
\newcommand\en{\end{equation}}
\newcommand\cm{\rm cm}
\newcommand\kms{\rm{\, km \, s^{-1}}}
\newcommand\K{\rm K}
\newcommand\pc{\rm pc}
\newcommand\kyr{\rm kyr}
\newcommand\erg{\rm erg}
\newcommand\cmm{\rm{cm^{-3}}}
\newcommand\etal{{\rm et al}.\ }
\newcommand\sd{\partial}
\newcommand{\cmark}{\ding{51}}
\newcommand{\xmark}{\ding{55}}
\begin{abstract}
The structure of protostellar cores can often be approximated by isothermal Bonnor-Ebert spheres (BES) which are stabilized by an external pressure. 
For the typical pressure of $10^4k_B\,\mathrm{K\,cm^{-3}}$ to $10^5k_B\,\mathrm{K\,cm^{-3}}$ found in molecular clouds, cores with masses below $1.5\,{\rm M_\odot}$ are stable against gravitational collapse.
In this paper, we analyze the efficiency of triggering a gravitational collapse by a nearby stellar wind, which represents an interesting scenario for triggered low-mass star formation. 
We derive analytically a new stability criterion for a BES compressed by a stellar wind, which depends on its initial nondimensional radius $\xi_{max}$. 
If the stability limit is violated the wind triggers a core collapse. 
Otherwise, the core is destroyed by the wind.  
We estimate its validity range to $2.5<\xi_{max}<4.2$ and confirm this in simulations with the SPH Code GADGET-3. 
The efficiency to trigger a gravitational collapse strongly decreases for $\xi_{max}<2.5$ since in this case destruction and acceleration of the whole sphere begin to dominate. 
We were unable to trigger a collapse for $\xi_{max}<2$, which leads to the conclusion that a stellar wind can move the smallest unstable stellar mass to $0.5\,\mathrm{M_\odot}$ and destabilizing even smaller cores would require an external pressure larger than $10^5k_B\,\mathrm{K\,cm^{-3}}$. 
For $\xi_{max}>4.2$ the expected wind strength according to our criterion is small enough so that the compression is slower than the sound speed of the BES and sound waves can be triggered. 
In this case our criterion underestimates somewhat the onset of collapse and detailed numerical analyses are required.
\end{abstract}


\section{Introduction}
A Bonnor-Ebert sphere  (BES, \cite{bonnar1956boyle,ebert1955verdichtung}) is an often-used theoretical model in simulations for protostellar cores, which is defined as an isothermal, spherically symmetric gas distribution with density $\rho(r)$ that is self-gravitating, in hydrostatic equilibrium and supported by the pressure of the ambient medium.
Especially attractive is its well-defined density profile (see figure \ref{fig:bonnerEbert}) as well as its stability behaviour under a uniform external pressure.
Despite the turbulent nature of the interstellar medium observations show that some molecular cloud cores are in hydrostatic equilibrium and that their density distributions follow a Bonnor-Ebert profile (\cite{alves2001internal}).\\
Using the equation of hydrostatic equilibrium one can derive a density distribution $\rho(r)$, which can be nondimensionalized such that it depends on the nondimensional radius $\xi$. In theoretical studies, one then assumes that there exists an external pressure $P_{ext}$ which stabilizes the BES and therefore the density profile is cut-off at a chosen $\xi_{max}$, where the internal pressure is $\rho(\xi_{max}) c_s^2 = P_{ext}$. Here $c_s$ is the isothermal sound speed of the core.
\cite{bonnar1956boyle} and \cite{ebert1955verdichtung} showed that the BES is unstable for $\xi_{max} > 6.45$, which means that perturbations can grow and the BES finally collapses, while for smaller radii it stays stable. 
Previous studies can be divided into two categories: On the one hand studies analyzing isolated BES and on the other hand studies analyzing BES which interact with other objects. The main focus of the first type is a better understanding of the evolution of the collapse of BES \citep{hunter1977collapse, ogino1999gravitational, banerjee2004formation,keto2010dynamics} or the derivation of new stability criteria for modified BES such as nonisothermal BES (\cite{sipila2011stability,sipila2015stability,sipila2017stability, nejad2016modified}).
The second type focuses on BES growth and induced star formation e.g. by mergers (\cite{burkert2009inevitable}).\\
Many studies showed that radiation from nearby stars can ionize the outer layers of a BES which leads to an additional pressure force that can trigger the collapse of the BES or disperse it \citep{kessel2003radiation,gritschneder2009ivine, bisbas2011radiation, ngoumou2014first, krumholz2014star}. 
Flow driven triggered star formation is another mechanism which can induce the collapse of a BES \citep{frank2015triggered}. Here gas flows from distant supernovae or stellar winds collide with the cloud core and act as an additional ram pressure source. The likelihood of collapse and star formation depends on the Mach number of the flow since higher Mach numbers can lead to instabilities in the BES which can even disperse the original cloud \citep{boss2008simultaneous, boss2013triggering,dugan2017agn}.
\cite{ngoumou2014first} analyzed the combined effect of stellar winds and ionizing radiation on a BES. They concentrated on the regime of low Mach number stellar winds, which means that the flow can be approximated by an additional external analytic pressure and the flow itself does not have to be simulated directly. 
In their simulations, which did not include a stabilizing pressure of the ambient medium, the wind itself was not able to trigger a collapse of the BES and they argued that ionization is the main force behind triggering star formation.\\
In this paper, we use the stellar wind model of \cite{ngoumou2014first} and show, that the wind alone can in fact trigger a collapse if we add the stabilizing external pressure from the ambient medium.
We assume that the wind is generated by a star at a distance $d$ with a mass-loss rate $\dot{M}$ and a wind velocity $v_w$.
In section \ref{sec:analyticDerivcation} we derive analytically a stability criterion for the BES which only depends on these parameters as well as on the radius $R$ of the BES, its isothermal sound speed $c_s$ and its initial nondimensional radius $\xi_{max}$. We assume here $d \gg R$, which means that the wind can be approximated to be parallel. We also estimate the range of $\xi_{max}$ for which this criterion should work. In section \ref{sec:simulations} we present simulations with the SPH-Code GADGET-3 \citep{springel2005cosmological} in which we analyze the stability of BES for different $\xi_{max}$ and stellar properties. We find here a good agreement with the analytical criterion derived in the previous section. 
In section \ref{sec:discussion} we discuss the implications of our results on the stability of protostellar cores in molecular clouds.

\section{Analytic derivation of the stability criterion}
\label{sec:analyticDerivcation}
\subsection{Definition of the Bonnor-Ebert sphere}
\begin{figure}[h]
\epsscale{0.8}
\plottwo{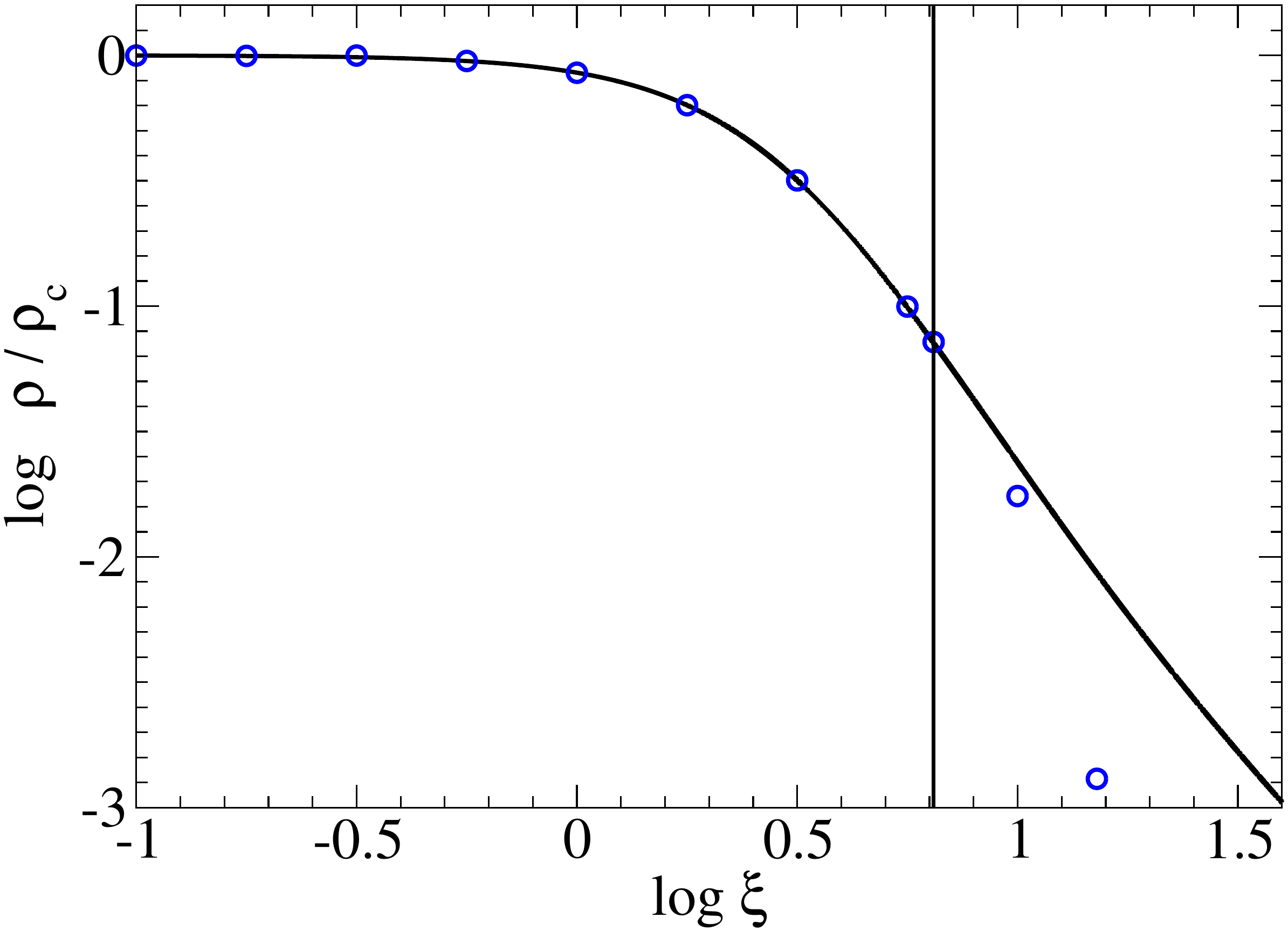}{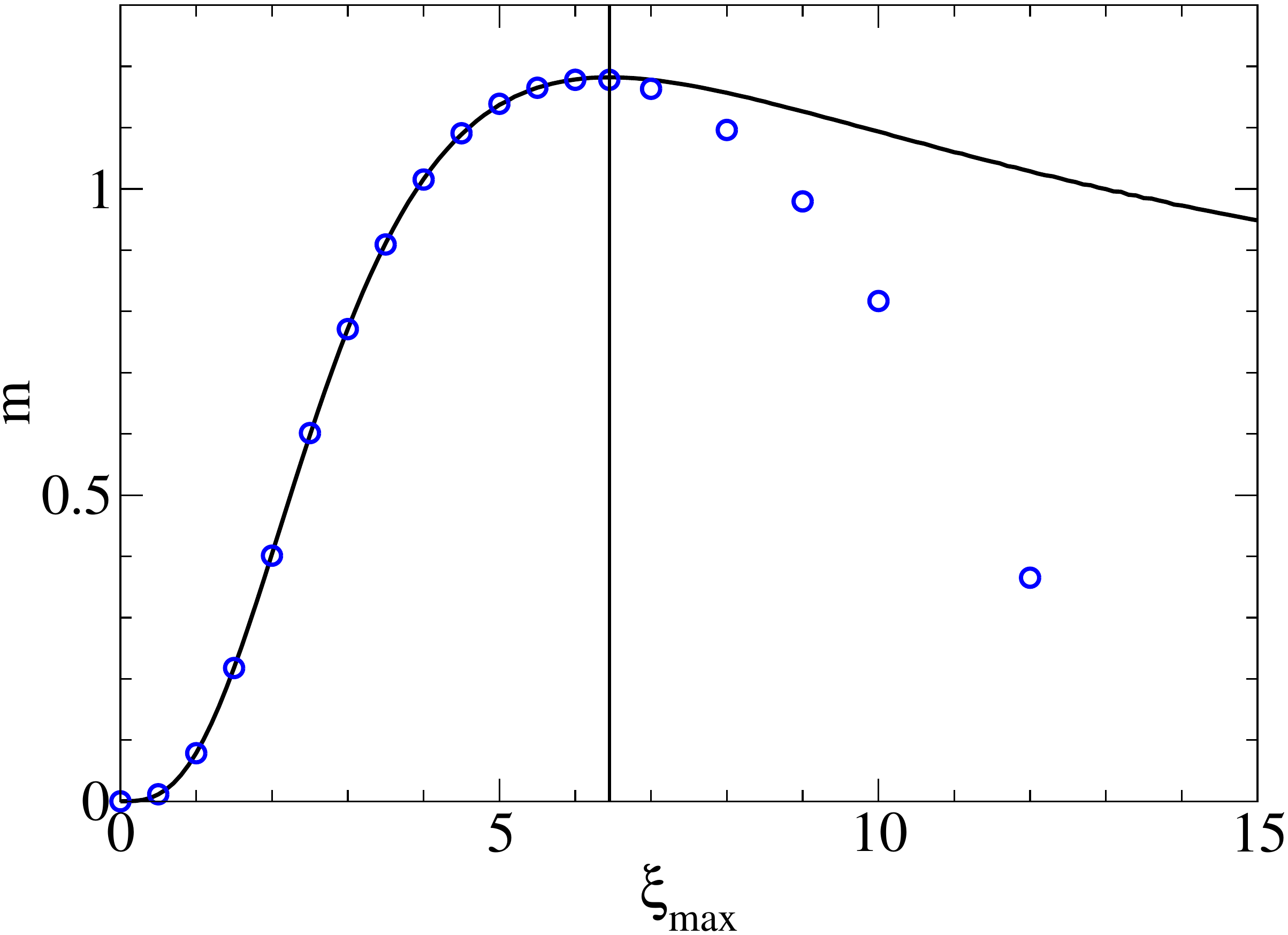}
\caption{Density distribution normalized to the central density $\rho_c$ of an isothermal Bonner-Ebert sphere and the total dimensionless mass of a Bonner-Ebert sphere as a function of its dimensionless radius $\xi$.
The blue dots show the approximations (\ref{eq:approxDensity}) and (\ref{eq:approxMass}). The vertical lines show the boundary  between the stable and unstable region.}
\label{fig:bonnerEbert}
\end{figure}
To fufill the properties described above, the BES has to fulfill the equation of hydrostatic equilibrium, Poisson's equation and an isothermal equation of state:
\begin{subequations}
\begin{equation}
\frac{1}{\rho}\nabla P = -\nabla \Phi,
\end{equation}
\begin{equation}
\nabla^2 \Phi = 4 \pi G \rho,
\label{eq:poissionBonnorEbert}
\end{equation}
\begin{equation}
P = \rho c_s^2.
\end{equation}
\end{subequations}
In the case of spherical symmetry the first and third equation lead to
\begin{equation}
\rho(r) = \rho_c \exp\left(\frac{-\Phi(r)}{c_s^2}\right)
\end{equation}
with the central density $\rho_c$ and the definition $\Phi(0) = 0$ for the gravitational potential.
By plugging this into equation (\ref{eq:poissionBonnorEbert}) we find:
\begin{equation}
\frac{1}{r^2}\frac{d}{dr}\left(r^2 \frac{d\Phi}{dr}\right) = 4 \pi G \exp\left(- \frac{\Phi(r)}{c_s^2}\right).
\label{eq:BonnorEbertSphereEquation}
\end{equation}
By introducing the dimensionless variables $\psi = \Phi/c_s^2$ and $\xi = \left(4 \pi G \rho_c / c_s^2\right)^{1/2} r$ we can simplify equation  (\ref{eq:BonnorEbertSphereEquation}) to a special form of the Lane-Emden equation
\begin{equation}
\frac{1}{\xi^2} \frac{d}{d\xi} \left(\xi^2 \frac{d\psi}{d\xi}\right) = \exp\left(-\psi\right),
\end{equation}
which can be solved numerically with the   boundary conditions $\psi(\xi = 0)= 0$ and $\left(d\psi /d\xi\right)_{\xi = 0} = 0$. In figure \ref{fig:bonnerEbert} we show the result $\rho(r) / \rho_c = \exp(-\psi(r))$ that can be well approximated within the stable region by:
\begin{equation}
\frac{\rho}  {\rho_c} ( \xi) = e^{-0.8444 \xi}  \left(1 + 0.8477 \xi + 0.1961 \xi^2 - 0.07308 \xi^3 + 0.01252 \xi^4\right).
\label{eq:approxDensity}
\end{equation}
The BES is typically confined by an external pressure $P_{ext}$ which determines the radius $R$ of the sphere by the condition $\rho(R) c_s^2 = P_{ext}$:
\begin{equation}
R = \left(\frac{c_s^2}{4 \pi G \rho_c}\right)^{1/2} \xi_{max}
\label{eq:BESFormulaR}
\end{equation}
with $\psi(\xi_{max}) = \psi_{max}= -\ln\left(P_{ext}/\left(\rho_c c_s^2\right)\right)$.
The total mass of the sphere is
\begin{equation}
M(\xi_{max}) = 4 \pi \int_0^R  r'^2 \rho(r') dr' = 4 \rho_c^{-1/2} \left(\frac{c_s^2}{4\pi G}\right)^{3/2} \left(\xi^2 \frac{d\psi}{d\xi}\right)_{\xi =\xi_{max}}
\end{equation}
and can be made dimensionless by the definition
\begin{equation}
m(\xi_{max}) = \frac{P_{ext}^{1/2} G^{3/2} M}{c_s^4} = \left(4 \pi \frac{\rho_c}{\rho(R)}\right)^{-1/2} \left(\xi^2 \frac{d \psi}{d\xi}\right)_{\xi = \xi_{max}}.
\label{eq:bonnorEbertAnhang}
\end{equation}
$m(\xi)$ is shown in the right panel in figure \ref{fig:bonnerEbert} and can be approximated by:
\begin{equation}
m(\xi) = \left\{
\begin{array}{ll}
-0.00131 \xi^2 + 0.109 \xi^3 - 0.0291 \xi^4 & 0 \leq \xi \leq 2 \\
e^{0.1536 (\xi-2)} \left(0.4012 + 0.3678 (\xi-2) - 0.1178 (\xi-2)^2 + 
   0.01015 (\xi-2)^3\right) & 2 \leq \xi \leq 5.5 \\
 1.16503 + 0.04056 (\xi-5.5) - 0.02852 (\xi-5.5)^2 + 
 0.005152 (\xi-5.5)^3 & 5.5 \leq \xi \leq 6.45 \\  
\end{array}
\right.
\label{eq:approxMass}
\end{equation}
 It reaches a maximum $m_{max} = 1.18$ for $\xi_{max} = 6.45$ which corresponds to a density contrast $\rho_c / \rho(R) = 14$ and all configurations with $\xi_{max} > 6.45$ lead to a gravitational collapse.

\subsection{Derivation of the stability criterion}
The starting point for our analysis is a BES in equilibrium which is stabilized by an external pressure
\begin{equation}
    P_{ext} = \frac{c_s^8}{G^3} \frac{m^2}{M^2}.
    \label{eq:pextBES}
\end{equation}
To first approximation one could add an uniform pressure $P_W$ representing the effect of the wind. In this case,  $m$ can be increased until the BES becomes unstable for $m > m_{max} = 1.18$. The critical uniform $P_W$ for a BES with initially $m=m_0$ is therefore given by:
\begin{equation}
    P_{W, crit} = P_{ext} \left( \frac{m_{max}^2}{m_0^2}-1\right).
    \label{eq:stabilityCriteriaBES}
\end{equation} 

\begin{figure}[ht]
\begin{center}
\epsscale{0.4}
\plotone{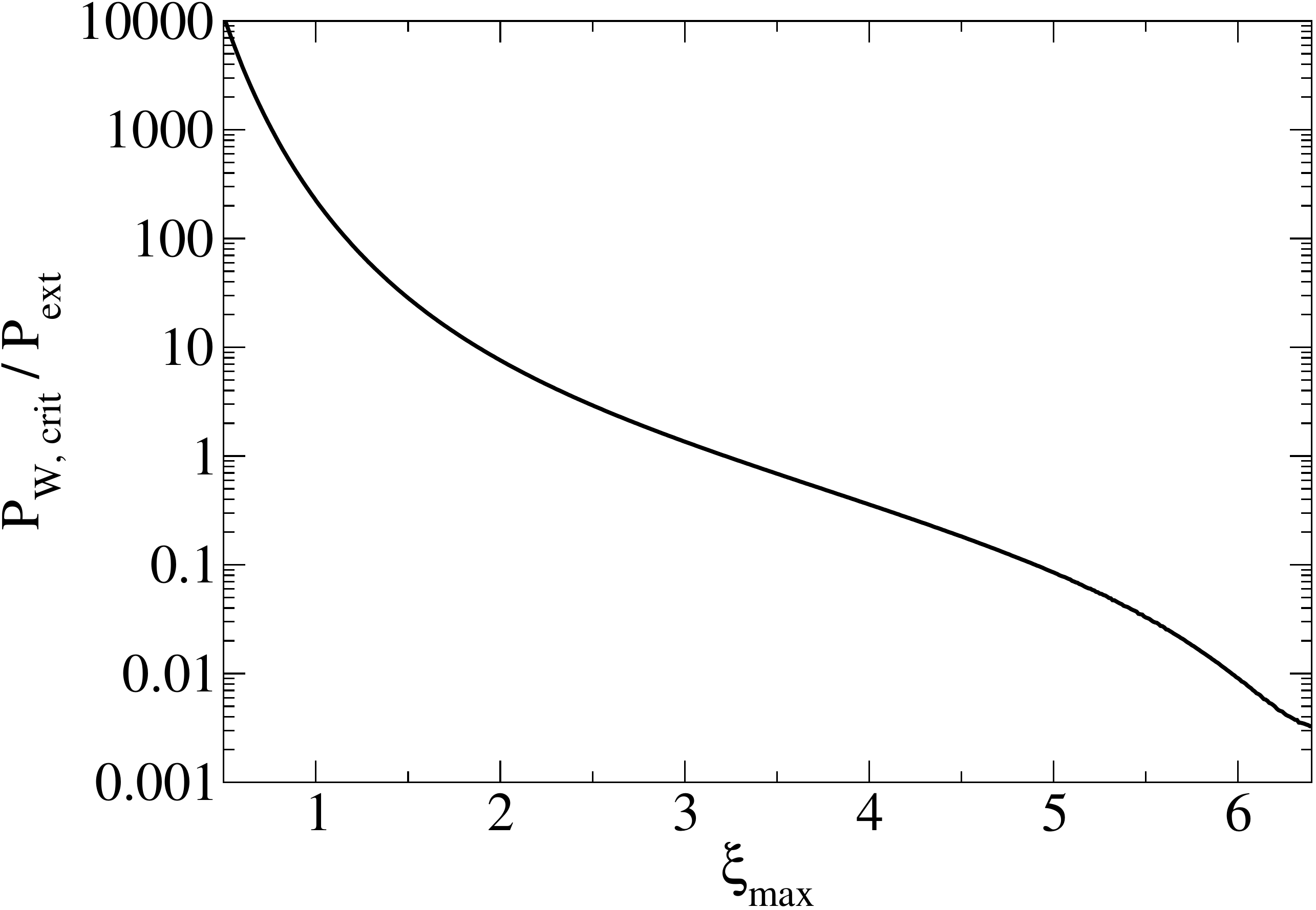}
\caption{The ratio of the critical external pressure from the wind to the initial external pressure as a function of the non-dimensional radius $\xi_{max}$ (see equation (\ref{eq:stabilityCriteriaBES})).}
\label{fig:bonnoerEbertstabilityPressure}
\end{center}
\end{figure}

However, the parallel wind impacts the clump from one side. It therefore does not lead to an isotropic pressure and the pressure $P_W$ that is required to trigger a collapse cannot directly be calculated but has to be approximated. We use the ansatz:
\begin{equation}
P_W = C \frac{\dot{M} v_w}{4\pi d^2}
\end{equation}
where C is a nondimensional parameter which should typically be O(1). $C=1$ corresponds to the maximum wind pressure at any point of the BES, $C = 1/3$ to the wind pressure averaged over the half of the surface of the BES that is directly affected by the wind and $C = 1/6$ to the wind pressure averaged over the whole surface. 
We now can  rewrite the original external pressure term using equation (\ref{eq:BESFormulaR}) and the density $\rho_B$ at the surface of the BES:
\begin{equation}
P_{ext} = c_s^2 \rho_B = c_s^2 \rho_c \frac{\rho_B}{\rho_c} = \frac{c_s^4 \xi_{max}^2}{4\pi G R^2} \frac{\rho_B}{\rho_c}.
\end{equation}
Plugging this into equation (\ref{eq:stabilityCriteriaBES}) we finally find for the condition that the BES is unstable:
\begin{equation}
\frac{G \dot{M} v_w R^2}{d^2 c_s^4} > \frac{\xi_{max}^2}{C} \frac{\rho_B}{\rho_c} \left( \frac{m_{max}^2}{m_0^2}-1\right).
\label{eq:StabilityCriteriaPressure}
\end{equation}
The right hand of this inequality is only a function of $\xi_{max}$ and C and can easily be calculated.
Figure \ref{fig:BESStability} shows the dependence of the left side of $\xi_{max}$.\\

\subsection{Expected validity range of the stability criterion}
\label{sec:validyRange}
In the derivation of inequality  (\ref{eq:StabilityCriteriaPressure}) we used several approximations, which we hide in the parameter C.
We assumed that we still can use the equations (\ref{eq:pextBES}) and (\ref{eq:stabilityCriteriaBES}) for a uniform pressure, which means the external wind pressure is only a perturbation to the hydrostatic equilibrium of the BES: This does not hold anymore for the case  $P_{ext} \ll P_W$ and simulations show deviations from inequality (\ref{eq:StabilityCriteriaPressure}) for $\xi_{max} < 2.5$ (see section \ref{subsec:results}).\\
A one-sided pressure component can also accelerate substantially the BES as a whole downstream if the rate of compression of the BES is smaller than its sound speed and sound waves are triggered. 
To derive the $\xi_{max}$ range in which this is the case we will  assume a pressure equilibrium between the internal and external pressure and take $P_{W}$ as a perturbation. 
For a BES in equilibrium, the last assumption is at least true at the beginning of the simulation. 
We assume the wind to be parallel to the x-axis, the centre of the BES lying on the axis, the point of the surface hit first by the wind to be at the origin of the coordinate system and we concentrate on the evolution of the BES on the x-axis.
We then find from Newton's law:
\begin{equation}
\frac{\dot{M} v_w}{4\pi d^2} = P_{C1} =  \rho_0 \frac{d}{dt} \left(x \dot{x}\right) = \rho_0 \frac{d^2}{dt^2} \left(\frac{x^2}{2}\right),
\end{equation}
where $\rho_0$ is the mean clump density and $x(t)$ is the position of the point upstream to the wind as function of time t.
With the initial conditions $x(0) = 0$ and $\dot{x} (0)= 0$ we find the solution for $t > 0$ (note there is a discontiuity in $\dot{x}$ at $t=0$):
\begin{equation}
x(t) = \sqrt{\frac{ P_{C1}}{\rho_0}} t; v_{compr}= \dot{x} = \sqrt{\frac{P_{C1}}{\rho_0}}.
\end{equation}
We expect a non-negligible fraction of the imposed wind momentum to go directly into acceleration of the whole sphere for $v_{compr} / c_s < 1$ 
as in this case the whole core can react to the momentum transfer by the wind, impacting from the left. With the definition $\rho_0 = B(\xi_{max})\rho_c$ we find that acceleration effects could matter for: 
\begin{equation}
\frac{G \dot{M} v_w R^2}{d^2 c_s^4} < \xi_{max}^2 B(\xi_{max}).
\label{eq:stabilityCriteriaCs}
\end{equation}
The right hand is again a function of $\xi_{max}$ and if we choose $B= \rho_B /\rho_C$ and $C= 1/3.8$ (see section \ref{subsec:results}) we expect for $\xi_{max} \geq 4.2$ deviations from inequality (\ref{eq:StabilityCriteriaPressure}).
It is important to stress that the inequality (\ref{eq:stabilityCriteriaCs}) should not be interpreted as a stability criterion on its own since we highly simplified the evolution of the compression of the BES and a gravitational collapse can also occur if there are sound waves. Vice versa even a collapsing core will have been accelerated as a whole to some extend. 
Especially for a BES that is barely stable a small compression is sufficient to deepen the gravitational potential enough in order to trigger a collapse that is much faster than a sound wave.
In summary we expect inequality (\ref{eq:StabilityCriteriaPressure}) to be valid for:
\begin{equation}
2.5 < \xi_{max} < 4.2.
\end{equation}

\section{Simulations}
\label{sec:simulations}
\subsection{Numerical methods}
For all simulations presented in this paper we used an improved version of the SPH code GADGET-3 \citep{springel2005cosmological} which is presented in detail in \cite{beck2015improved}.
It supports the density formulation of SPH with entropy as the thermodynamical variable
as in \cite{springel2002cosmological}, a time-dependent artificial viscosity according to \cite{cullen2010inviscid} as well as adaptive gravitational softening as presented in \cite{iannuzzi2011adaptive}, which is important to achieve hydrostatic equilibrium.
We also use the wakeup scheme presented in \cite{pakmor2012stellar}, which activates particles if there are other particles in the smoothing kernel with much smaller timesteps.
In all simulations, we use the Wendland $C^4$ kernel \citep{dehnen2012improving} with 200 neighbours in the SPH kernel.
For the equation of state we use
\begin{equation}
P = \mathrm{max} \left( c_s^2 \rho \left[1 +\left(\frac{\rho}{\rho_{crit}}\right)^{\gamma-1}\right] -P_{ext},0\right)
\label{eq:eos}
\end{equation}
with $\mathrm {\rho_{crit} = 10^{-13} \,g\,cm^{-3}}$, $c_s = 200\mathrm{\,m\,s^{-1}}$ corresponding to $T=10\,\mathrm{K}$ and the external pressure $P_{ext}$ to stabilize the BES. The idea behind equation (\ref{eq:eos}) is that for $\rho \ll \rho_{crit}$ the core can cool efficiently and is  isothermal.
 For $\rho \gg \rho_{crit}$ it becomes optical thick and the equation of state becomes adiabatic.  We create sink particles above the density threshold of $10^{-11} \,\mathrm{\,g\,cm^{-3}}$ and run all simulations with self-gravity and advanced SPH but without periodic boundary conditions. In all simulations we use $2\times 10^6$ particles, which means the typical mass of a SPH particle is O($2 \times 10^{-6} \,\mathrm{M_\odot}$). The minimum Jeans mass for $\rho = \rho_{crit}$ and $T=10 $K is $M_{j} = 4\times 10^{-3} \,\mathrm{M_\odot}$ which corresponds to around 2000 particles. This is more than $2 N_{neigh} = 400$ for our kernel and can therefore always be resolved according to the criterion from \cite{bate1997resolution}.\\

For the wind we use a similar model as presented in \cite{ngoumou2014first}, which does not sample the wind ejecta itself but only takes into account the momentum of the wind ejecta:
We assume a constant mass-loss rate $\dot{M}$ and terminal wind velocity $v_w$, which is equivalent to a momentum production rate of $dp /dt = \dot{M} v_w$.
In every time step, the momentum generated since the last time step is injected isotropically in the environment of the star using the HEALPix algorithm \citep{gorski2005healpix}. 
If in one of the pixels no SPH particle is found, the corresponding momentum is ignored.
As in \cite{ngoumou2014first} we also allow the splitting of rays and smooth the momentum injection over one smoothing length. 
The splitting is especially important if $d \gg R$ holds since in this case the BES only spans a small solid angle seen from the star and most of the initial HEALPix rays are empty.

\subsection{Initial conditions}
\begin{figure} [hp]
\begin{center}
    \begin{minipage}{0.8\linewidth}
    \includegraphics[width=1\linewidth]{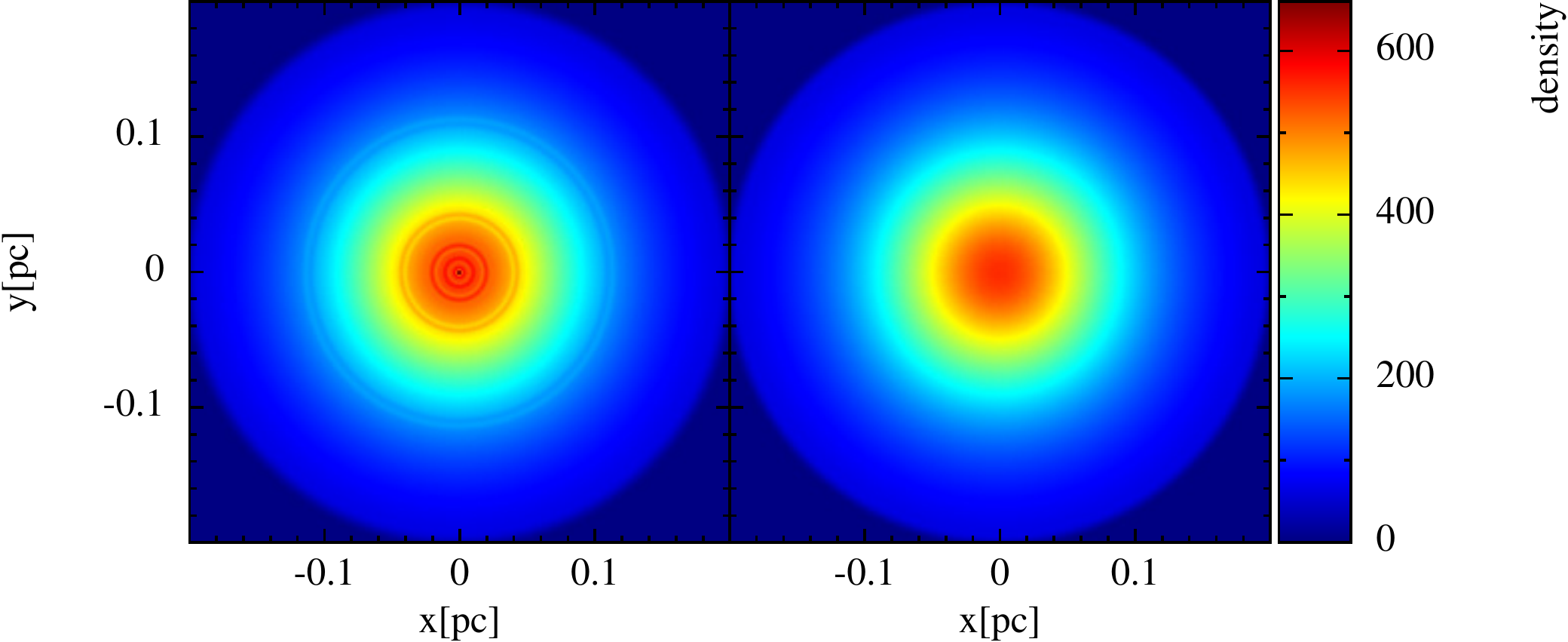}
    \end{minipage}\\
        \begin{minipage}{0.48\linewidth}
    \includegraphics[width=1\linewidth]{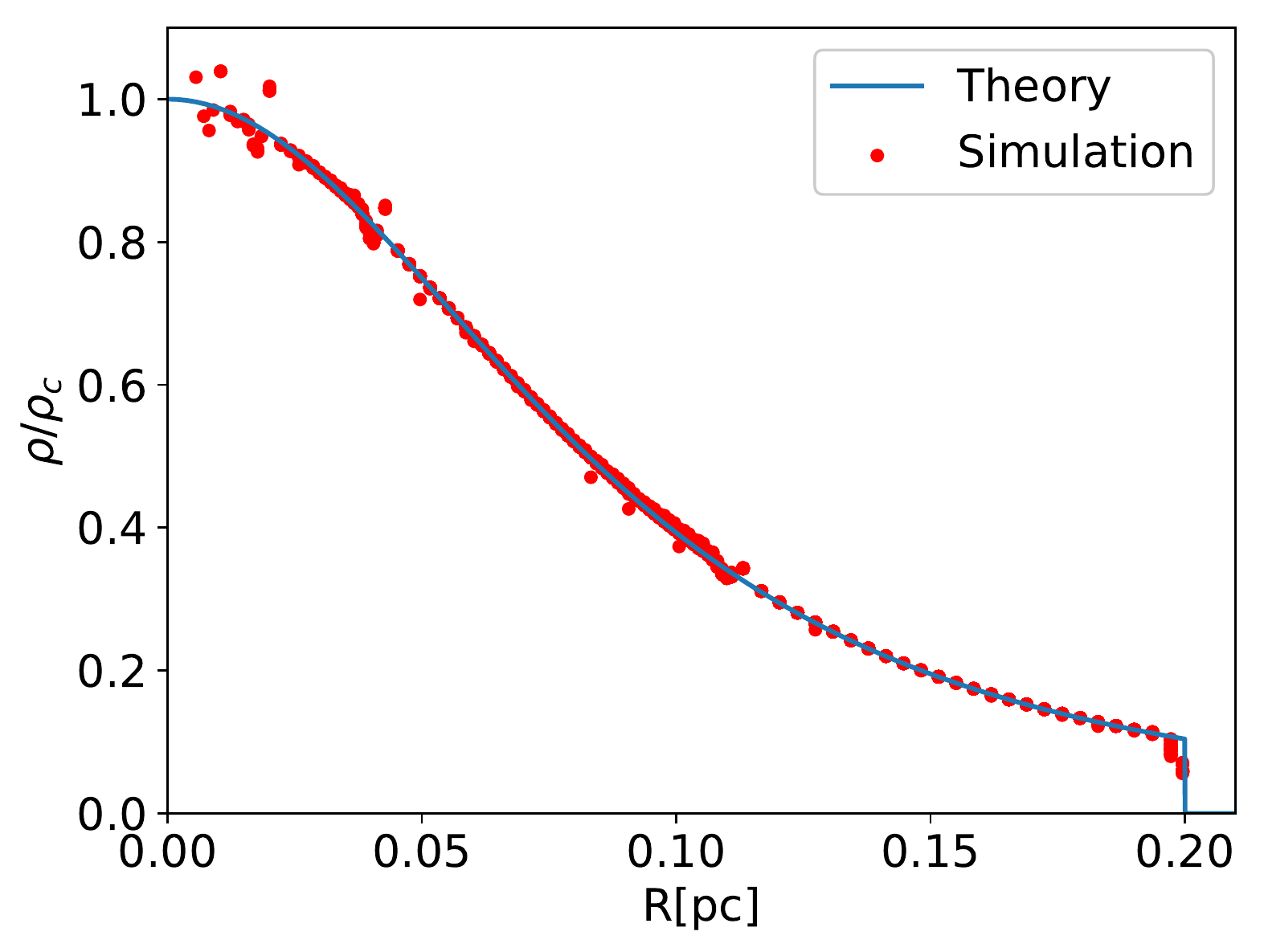}
    \end{minipage}
    \hfill
    \begin{minipage}{0.48\linewidth}
    \includegraphics[width=1\linewidth]{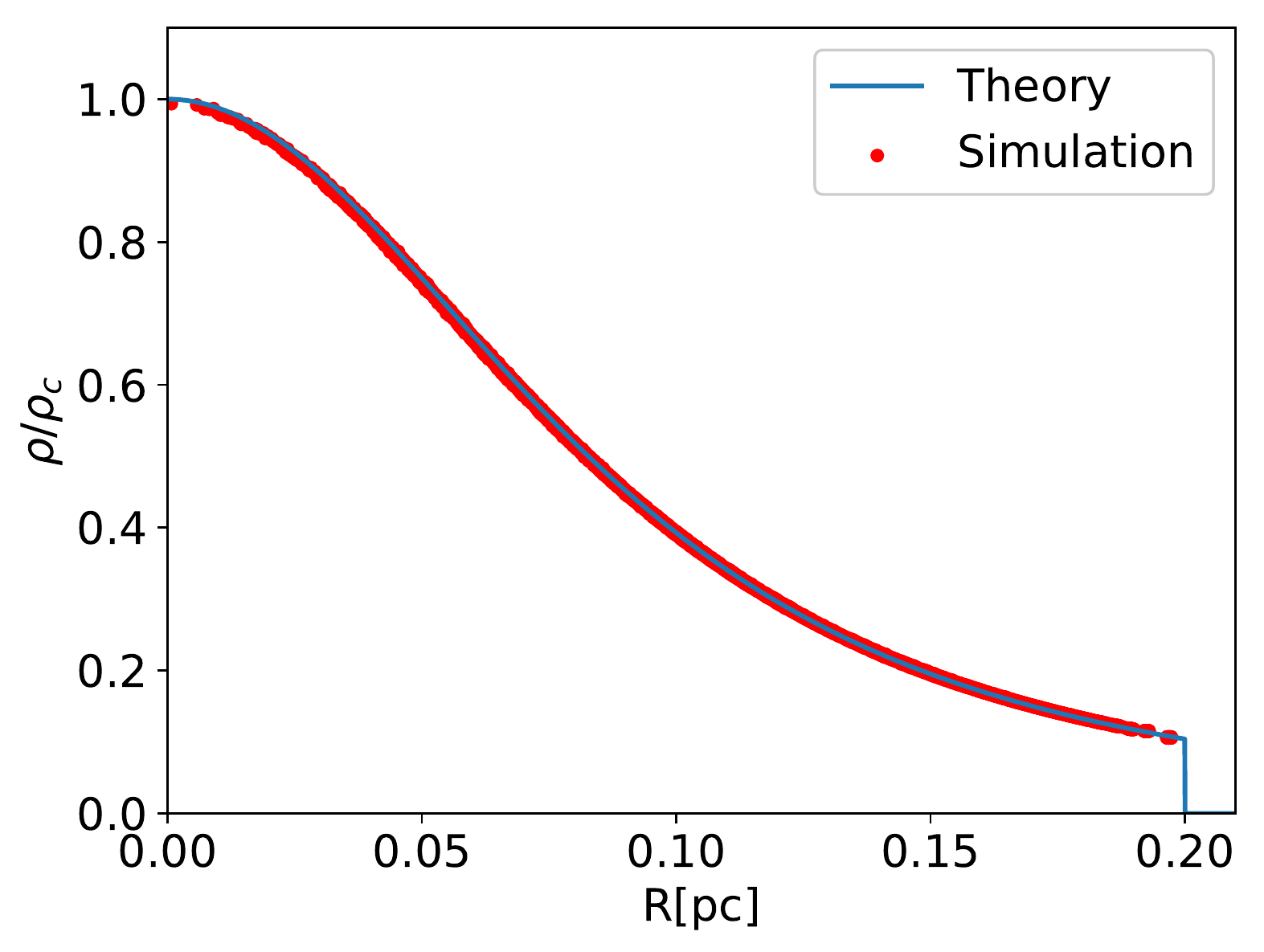}
\end{minipage}
\\
\end{center}
\caption{The left panel shows the initial conditions constructed with the method described in \cite{pakmor2012stellar} for a BES with $\xi_{max} = 5.5$ and radius $R=0.2\,\mathrm{pc}$. In the upper row we plot a density slice through the $z = 0$ plane (in $\mathrm{M_\odot \,pc^{-3} }$) and in the lower row we plot the normalized density for each 200th particle as a function of the radius. On the right hand we show the same properties after evolving the initial conditions from the left hand for $60\,\mathrm{Myr}$ under the influence of gravity and hydrodynamics.}
\label{fig:BESInitialConditions}
\end{figure}
Especially for $\xi_{max}$ close to $6.45$ it is important to reduce the noise in the initial conditions since too large noise can already trigger a collapse. We therefore use the method described in \cite{pakmor2012stellar}, that uses the HEALPix algorithm to create spherical symmetric initial conditions. The idea is to build up the sphere by spherical shells that themselves consist of approximately cubic cells, all with the same mass. In each cell one SPH particle is set. For a detailed mathematical description we refer to \cite{pakmor2012stellar}.
To further reduce the initial noise and also to get rid of preferred directions we let evolve the initial conditions without the wind until the density distribution becomes static. As one can see in the upper panel of figure \ref{fig:BESInitialConditions} the gaps between the different shells disappear with time.\\

 \subsection{Overview of simulations}
In all simulations we use  $R=0.2\,\mathrm{pc}$, $d=3\,\mathrm{pc}$ and fix $\xi_{max}$, $\dot{M}v_w$, $P_{ext}$ at the beginning of the simulation. 
Inspired by inequality (\ref{eq:StabilityCriteriaPressure}) we use the dimensionless physical quantity:
\begin{equation}
C_{sim} = 
 \frac{\xi_{max}^2 d^2 c_s^4}{G \dot{M} v_w R^2} \frac{\rho_B}{\rho_c} \left( \frac{m_{max}^2}{m_0^2}-1\right) = \frac{P_{W,crit}}{P_{C1}},
 \label{eq:definitionCSim}
\end{equation}
 together with $\xi_{max}$ to unambiguously classify the simulations. 
 We vary $C_{sim}$ and $\xi_{max}$ to find the boundary between the stable and unstable regime. All simulations are stopped after the formation of the first sink particle or when the sphere has dissolved (i.e. the maximum density is half of the maximum density at the beginning of the simulation).

\subsection{Results}
\label{subsec:results}
\begin{figure}[hp]
\begin{center}
    \begin{minipage}{0.7\linewidth}
    \includegraphics[width=1\linewidth]{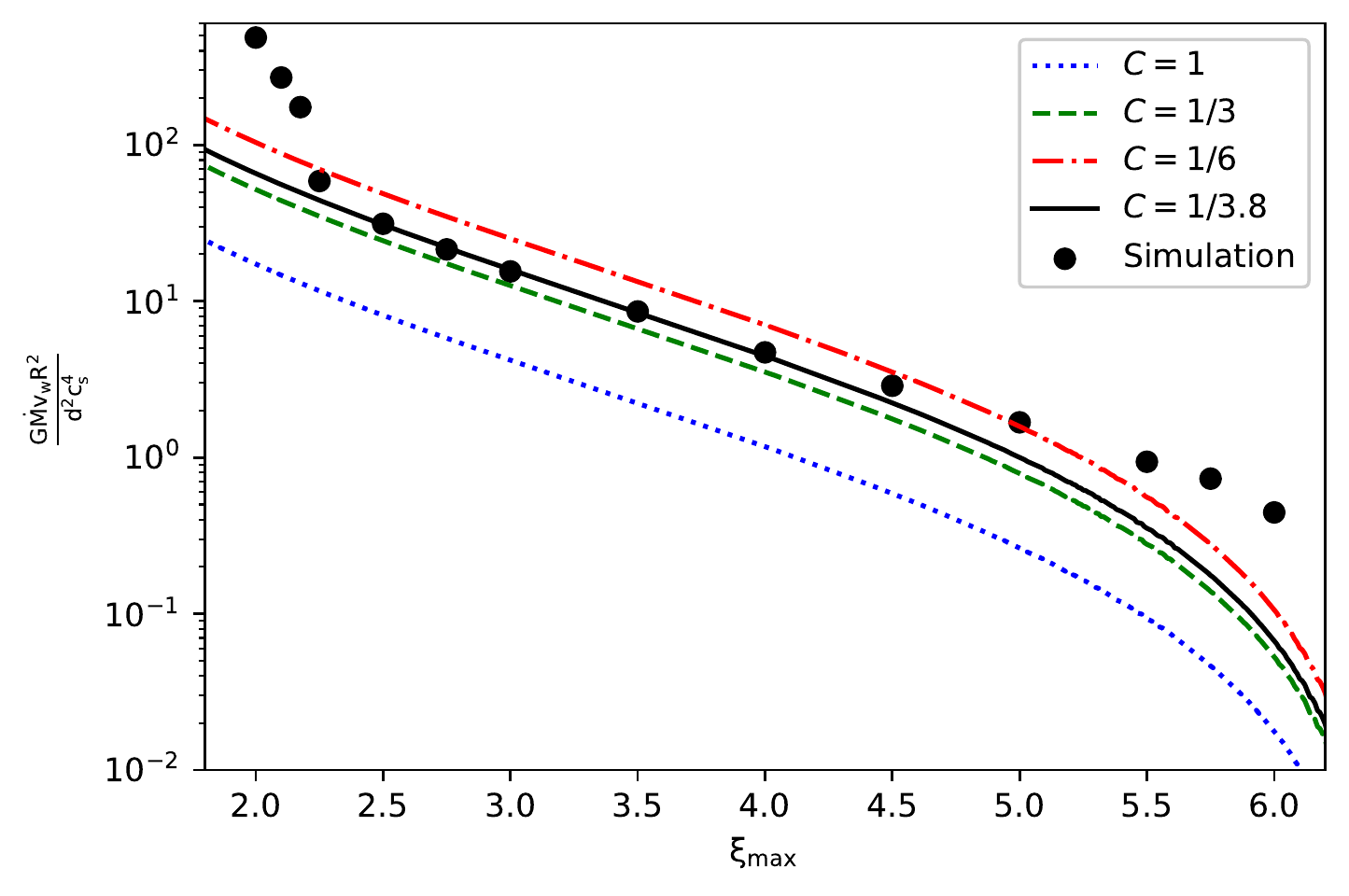}
    \end{minipage}\\
\end{center}
\caption{We show the boundary between the stable and unstable phase for the BES as a function of $\xi_{max}$.
The lines show the left side of inequality (\ref{eq:StabilityCriteriaPressure}) for different C and the black points represent simulations in which the BES is barely stable and does not collapse.
For $2.5 \leq \xi_{max} \leq 4$ the results from the simulations can be well described by $C = 1/3.8$.}
\label{fig:BESStability}
\end{figure}

\begin{figure}[hp]
\begin{center}
    \begin{minipage}{0.7\linewidth}
    \includegraphics[width=1\linewidth]{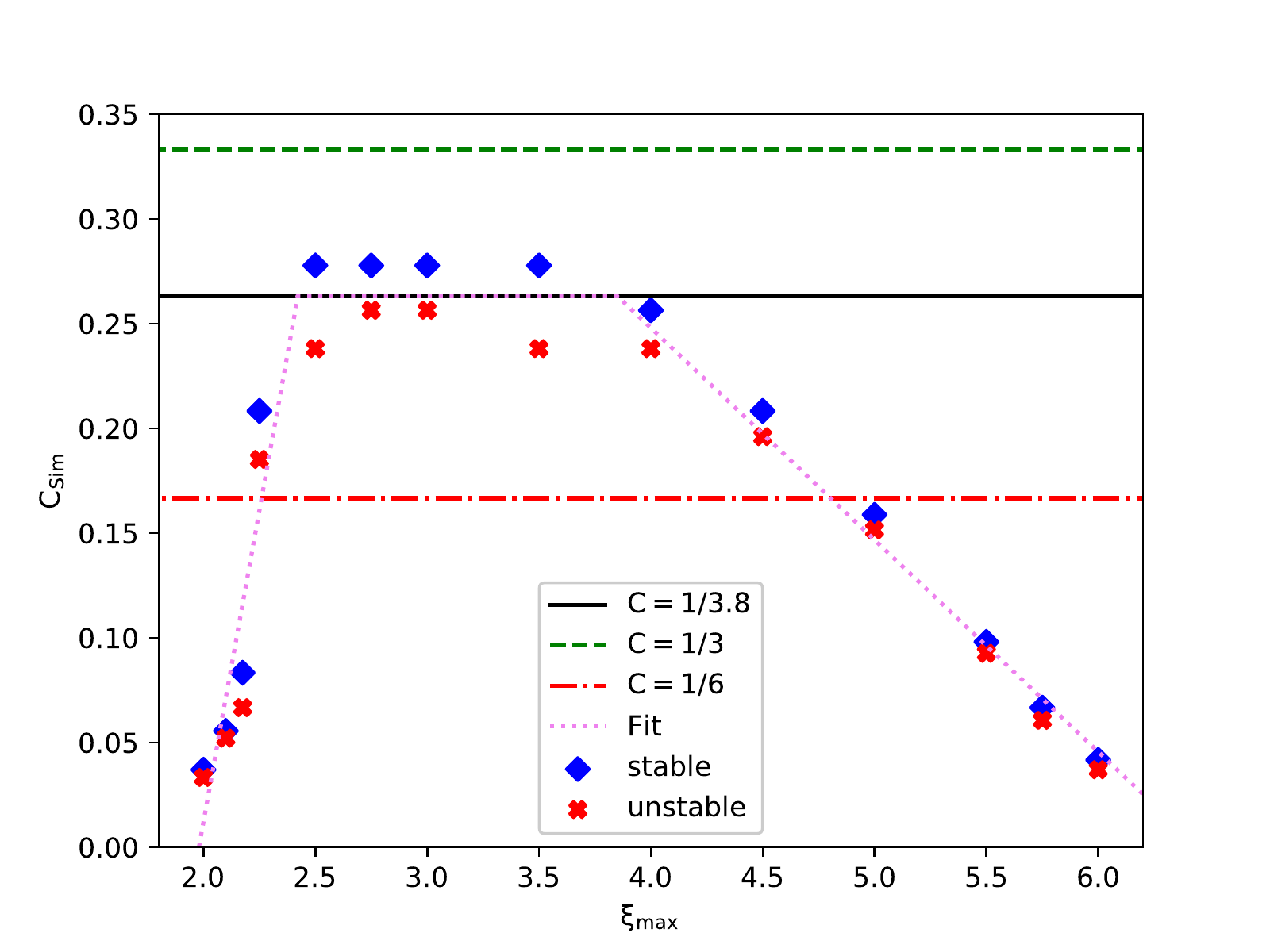}
    \end{minipage}\\
\end{center}
\caption{We show for simulations with different $C_{Sim}$ and $\xi_{max}$ whether a sink particle forms.
The dotted line shows the fitting function (\ref{eq:CsimFit}).}
\label{fig:BESStabilityC}
\end{figure}

As one can see in figure \ref{fig:BESStability} and \ref{fig:BESStabilityC} the stability of the BES can be well described by inequality (\ref{eq:StabilityCriteriaPressure}) with:
\begin{equation}
C = \left\{
\begin{array}{ll}
1/3.8 + (\xi-2.42) 0.595 &\, \mathrm{for}\, 2 \leq \xi \leq 2.42  \\
1/3.8 & \, \mathrm{for}\,2.42 \leq \xi \leq 3.8 \\
1/3.8 - (\xi-3.85) 0.1012 & \, \mathrm{for}\,3.85 \leq \xi \leq 6 \\
\end{array}
\right. 
\label{eq:CsimFit}
\end{equation}

\begin{figure}[hp]
\begin{center}
    \begin{minipage}{0.88\linewidth}
    \includegraphics[width=1\linewidth]{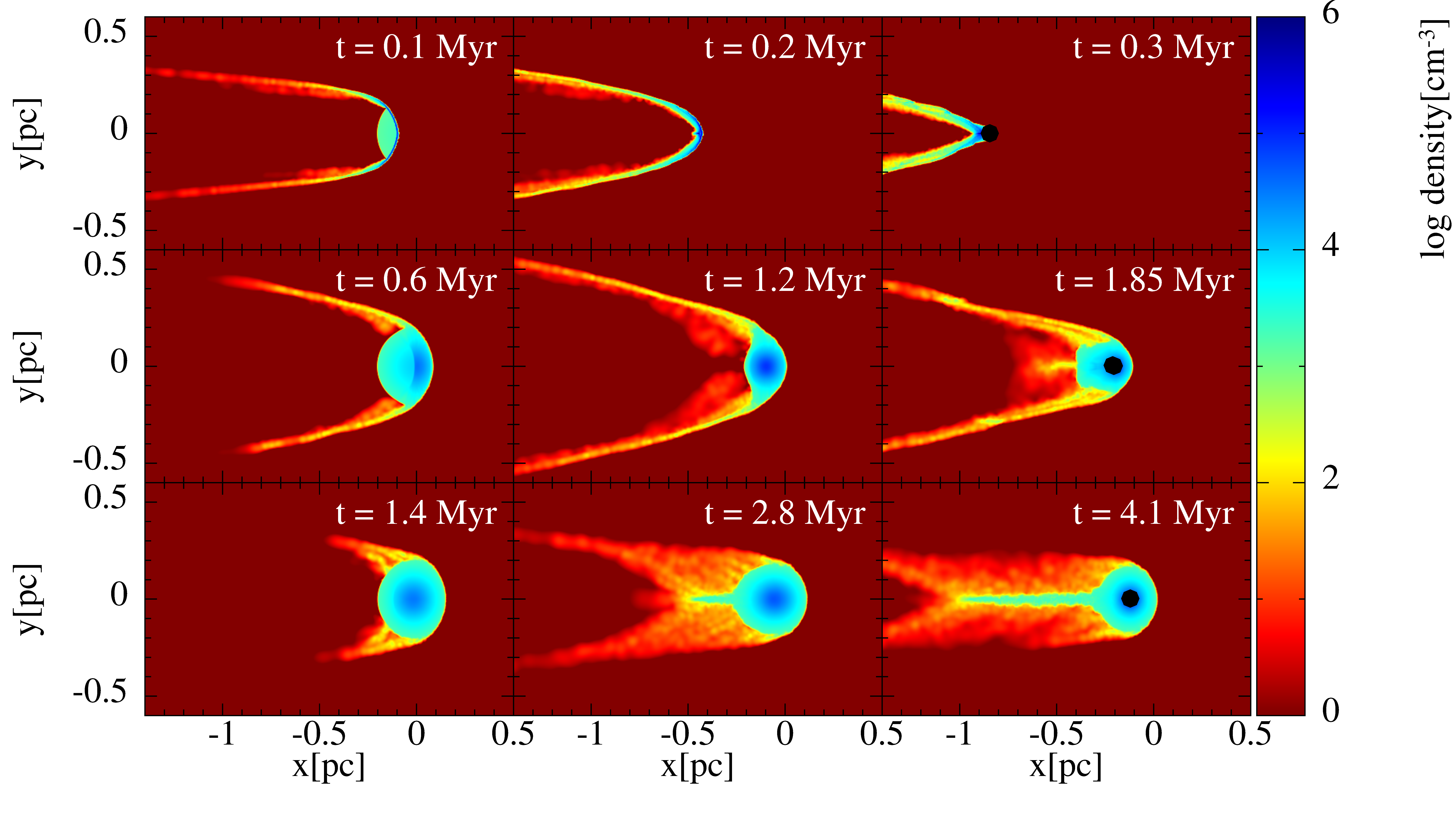}
    \end{minipage}\\
\end{center}
\caption{We show number density slices for the simulations with $C_{sim} = 0.03$ and $\xi_{max} = 2$ (top), $C_{sim} = 0.238$ and $\xi_{max} = 4$ (middle), $C_{sim} = 0.033$ and $\xi_{max} = 6$ (bottom). A sink particle is represented by a black circle.}
\label{fig:densitySplash}
\end{figure}

\begin{figure}[hp]
\begin{center}
    \begin{minipage}{0.86\linewidth}
    \includegraphics[width=1\linewidth]{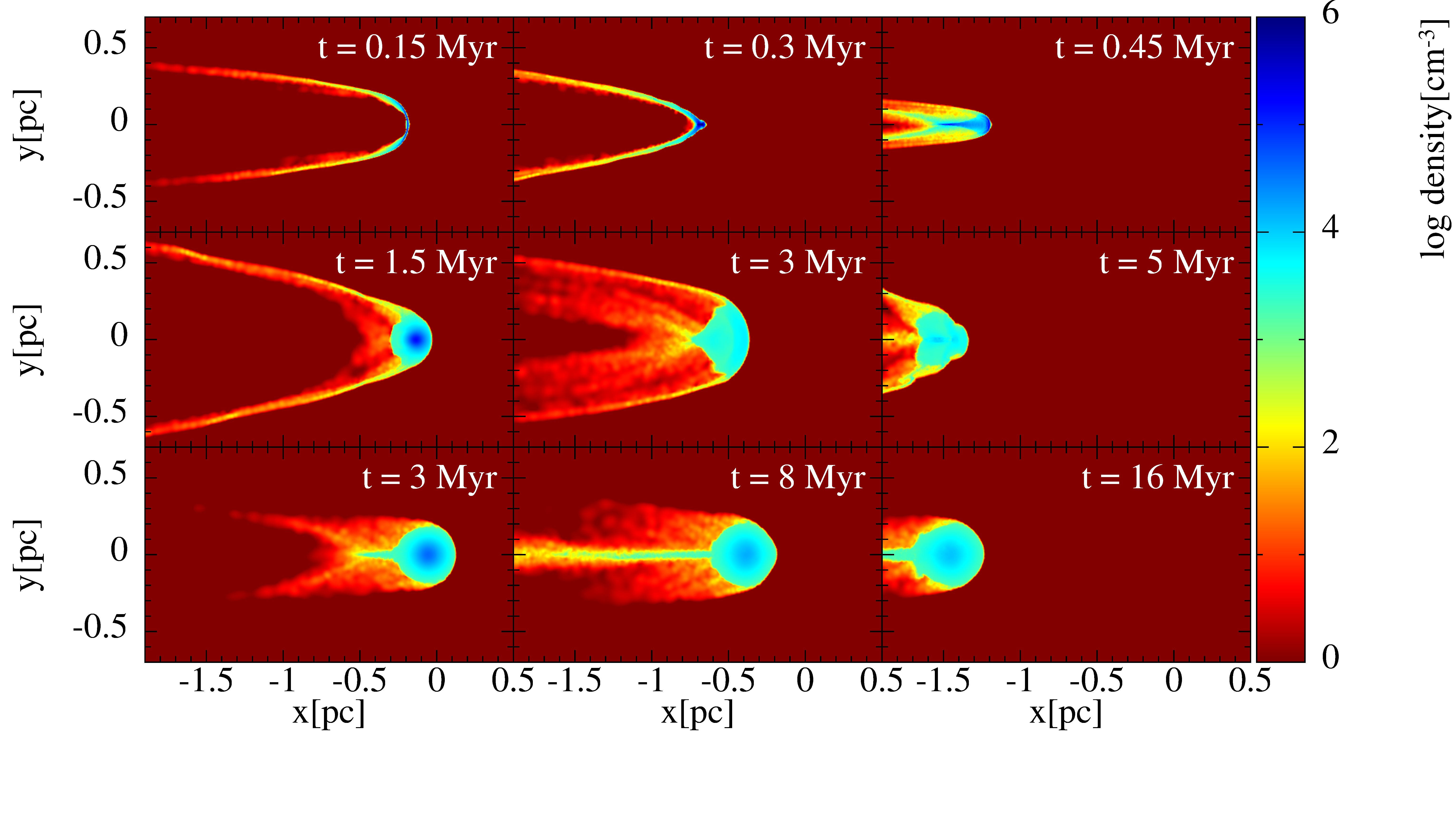}
    \end{minipage}\\
\end{center}
\caption{We show number density slices for the simulations with $C_{sim} = 0.042$ and $\xi_{max} = 2$ (top), $C_{sim} = 0.256$ and $\xi_{max} = 4$ (middle), $C_{sim} = 0.042$ and $\xi_{max} = 6$ (bottom). No sink particles are formed.}
\label{fig:DensityNoStar}
\end{figure}

\begin{figure}[hp]
\begin{center}
    \begin{minipage}{0.7\linewidth}
    \includegraphics[width=1\linewidth]{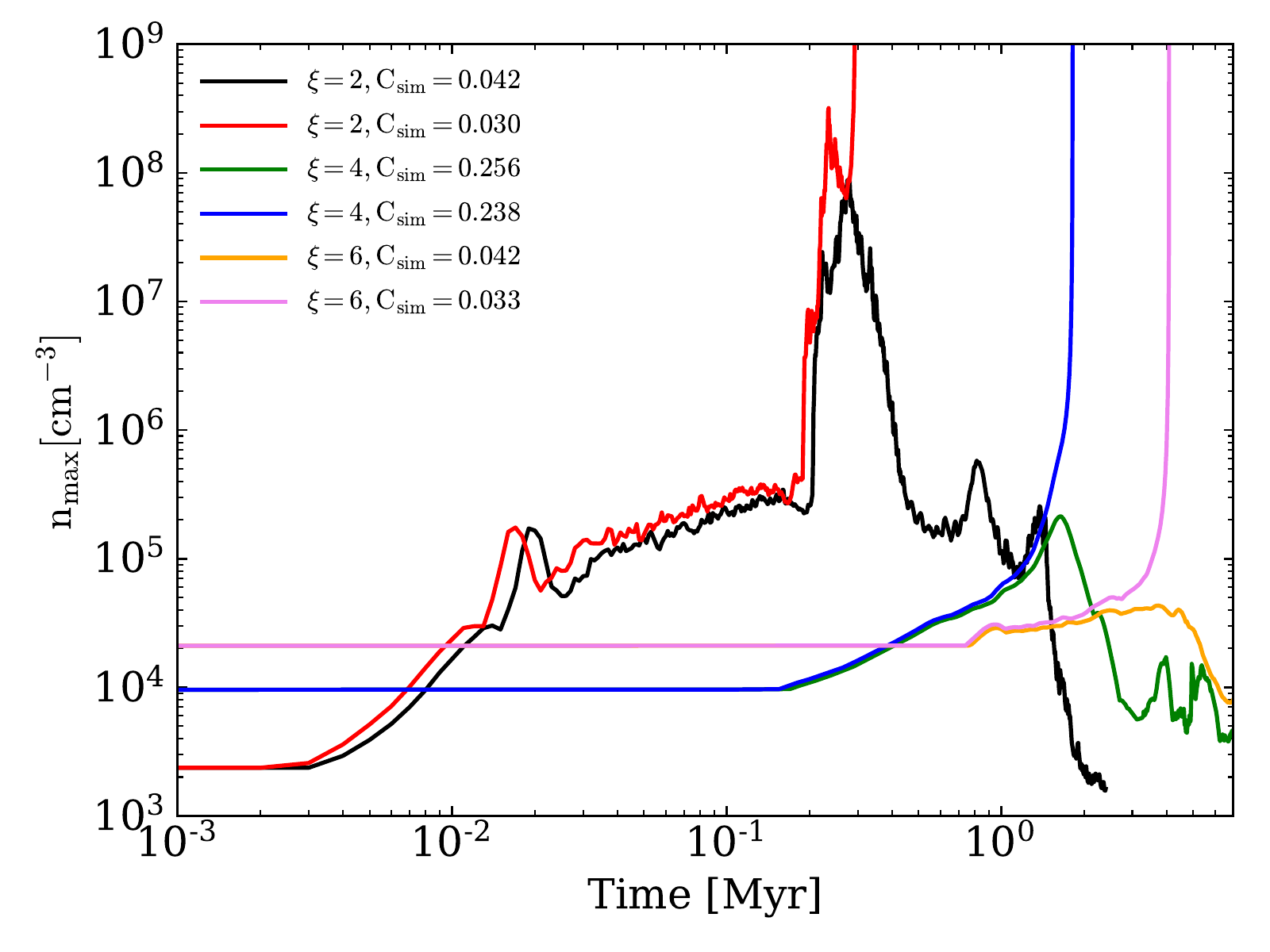}
    \end{minipage}\\
\end{center}
\caption{The evolution of the maximum density in simulations with different combinations of $C_{sim}$ and $\xi_{max}$.
In simulations in which sink particles form we can clearly see the runaway nature of the gravitational collapse.}
\label{fig:density_evolution}
\end{figure}

In figure \ref{fig:densitySplash} and \ref{fig:DensityNoStar} we show exemplarly the evolution of BES with $\xi_{max} =2$, $\xi_{max} =4$ and $\xi_{max} = 6$ and in figure \ref{fig:density_evolution} the evolution of their maximum density.
For $\xi_{max} =2 $ at the edge of the BES gas gets blown away while the rest gets compressed. The compression of the cloud is faster than the sound speed, which means the shielded side of the BES is not affected by the wind until the compression hits it. 
Afterwards, the whole remnant begins to move downstream. For $C_{sim} = 0.03$ at $t= 290\kyr$ a gravitational collapse is triggered. Due to the low mass of the remnant even a maximum density of $n = 10^{8} \,\mathrm{cm^{-3}}$ is not enough for $C_{sim} = 0.042$ to trigger a gravitational collapse. Clearly the assumption of spherical symmetry as made in section \ref{sec:analyticDerivcation} is violated.\\
For $\xi_{max} =4$ also gas at the surface of the BES is blown away and the compression is faster than the sound speed.
Now, the compression is enough to trigger a runaway gravitational collapse towards a star for $C_{sim} = 0.238$, while for $C_{sim}= 0.256$ the compression is too weak and the remnant of the BES gets blown away. 
For $\xi_{max} = 6$ in both simulations the compression is slower than the speed of sound which means a sound wave can reach the shielded surface of the BES before the compression front hits it. 
The sound wave leads to an expansion of the shielded side of the BES which prevents the gravitational collapse for $C_{sim} = 0.042$. For $C_{sim} = 0.033$ the compression is fast enough to trigger a gravitational collapse.

\section{Discussion}
\label{sec:discussion}
\begin{figure}[hp]
\begin{center}
    \begin{minipage}{0.53\linewidth}
    \includegraphics[width=1\linewidth]{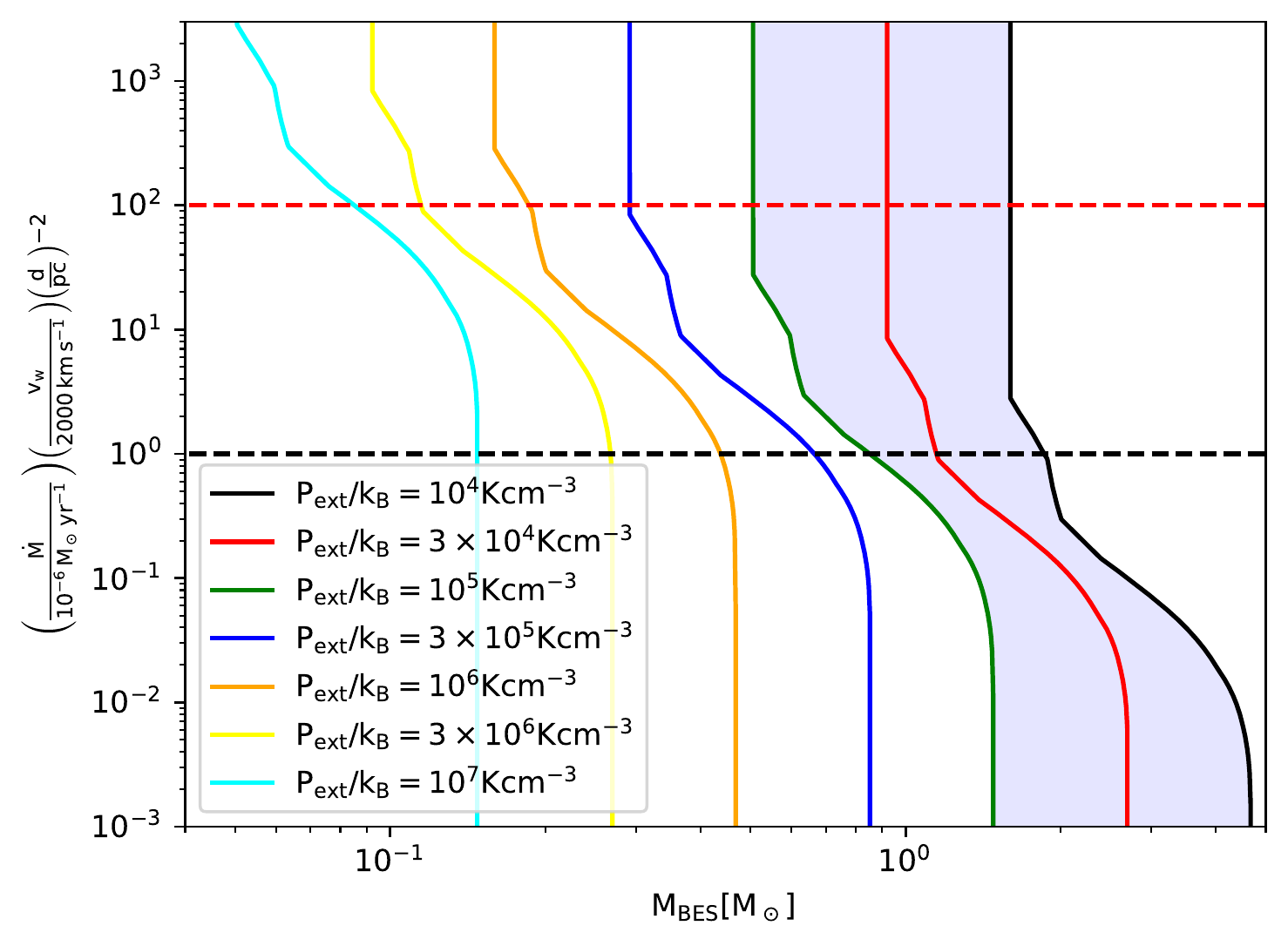}
    \end{minipage}\\
\end{center}
\caption{We show the critical wind strength required to trigger a gravitational collapse of a BES as a function of its mass for different external pressures $P_{ext}$ and $c_s = 200\mathrm{\, m \,s^{-1}}$.
We determine C by fitting it to the results presented in figure \ref{fig:BESStabilityC} and use $ C = 0$ for $\xi_{max} <2$.
The shaded area shows the typical external pressure found in molecular clouds,  the black dashed line the effect of a typical O star with distance $1\,\mathrm{pc}$ \added{and the red dashed line the effect of a strong Wolf-Rayet star with distance $1\,\mathrm{pc}$.}}
\label{fig:WindStrengthForPress}
\end{figure}

\begin{figure}[h]
\begin{center}
    \begin{minipage}{0.6\linewidth}
    \includegraphics[width=1\linewidth]{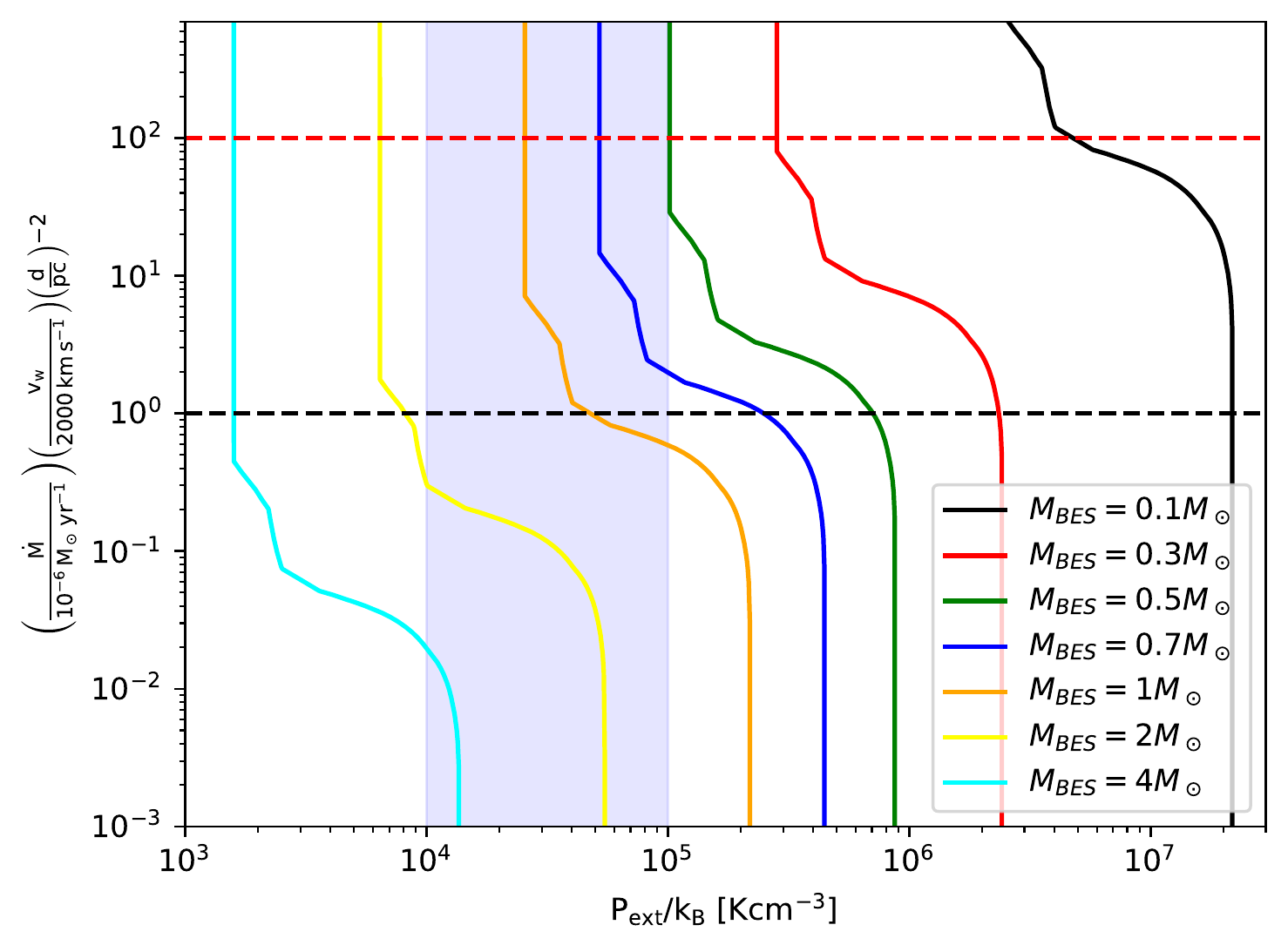}
    \end{minipage}\\
\end{center}
\caption{We show the critical wind strength required to trigger a gravitational collapse of a BES as a function of the external pressure for different masses $M_{BES}$ and $c_s = 200\mathrm{\, m \,s^{-1}}$.
We determine C by fitting it to the results presented in figure \ref{fig:BESStabilityC} and use $ C = 0$ for $\xi_{max} <2$.
The shaded area shows the typical external pressure found in molecular clouds,  the black dashed line the effect of a typical O star with distance $1\,\mathrm{pc}$ \added{and the red dashed line the effect of a strong Wolf-Rayet star with distance $1\,\mathrm{pc}$.}}
\label{fig:WindStrengthForM}
\end{figure}

In the previous sections we showed that a BES with $2<\xi < 6.45$ can be efficiently destabilized by a nearby wind source.
From equation (\ref{eq:bonnorEbertAnhang}) follows for the mass of the core:
\begin{equation}
M_{BES} = m(\xi_{max}) \frac{c_s^4}{G^{3/2} P_{ext}^{1/2}} \approx 3.96 \left(\frac{c_s}{200\,\mathrm{km\,s^{-1}}}\right)^4 \left(\frac{P_{ext}/k_B}{10^4\,\mathrm{K\, cm^{-3}}} \right)^{-1/2} m(\xi_{max}),
\end{equation}
which depends on the one hand on the inner properties of the BES (the sound speed) and on the other hand on the environment (the external pressure).
Although the pressure within molecular clouds can vary from $10^4 k_B\,\mathrm{K}\,\mathrm{ cm^{-3}}$ to $10^7 k_B \,\mathrm{K}\,\mathrm{ cm^{-3}}$ \citep{sun2020dynamical}, typical values from observations and simulations show a pressure between  $10^4 k_B\,\mathrm{K}\,\mathrm{ cm^{-3}}$ and $10^5 k_B\,\mathrm{K}\,\mathrm{ cm^{-3}}$ \citep{sun2020dynamical,heigl2018morphology,anathpindika2020filament} in the ISM.
As a result, cores with $M <1.5 \,\mathrm{M_\odot}$ are typically stable.
This boundary can be shifted to $0.5 \,\mathrm{M_\odot}$ by a nearby star  if we assume that that BES with $\xi =2$ can be destabilized. To further analyse the wind strength required to destabilize the BES one can rewrite inequality (\ref{eq:StabilityCriteriaPressure}) using equation (\ref{eq:BESFormulaR}):
\begin{equation}
\frac{\dot{M}v_w}{d^2}  > \frac{4\pi}{C} \left(1.18^2 \frac{c_s^8}{G^3 M_{BES}^2 }- P_{ext}\right).
\label{eq:criticalWind}
\end{equation}
In figure \ref{fig:WindStrengthForPress} and \ref{fig:WindStrengthForM} we show the right side of inequality (\ref{eq:criticalWind}) as a function of $M_{BES}$ and $P_{ext}$.
Given certain wind properties ($\dot{M}$, $v_w$, $d$) and a given external pressure, $P_{ext}$, collapse and star formation is triggered by the wind for BES that lie to the right of the corresponding colored line.
If one assumes, for example, a typical O star with $\dot{M} = 10^{-6} \mathrm{\,M_\odot \, yr^{-1}}$, $v_w =2000 \mathrm{\,km\,s^{-1}}$ at distance $1\,\mathrm{pc}$ it can trigger a collapse for small external pressures or large BES masses. 
If we follow in figure \ref{fig:WindStrengthForPress} the dashed, black line, we find that it has neglible effect on the stability for the orange line ($P_{ext} = 10^6 k_B\,\mathrm{K}\,\mathrm{ cm^{-3}}$) while for the black line ($P_{ext} = 10^4 k_B\,\mathrm{K}\,\mathrm{ cm^{-3}}$) the wind makes a substantial difference.
Similarly we find in figure \ref{fig:WindStrengthForM} that the wind  makes no substantial difference for a mass $M_{BES} < 0.5 \mathrm{M_\odot}$.
We also show exemplary in figure  \ref{fig:WindStrengthForPress} and \ref{fig:WindStrengthForM} the wind strength of a strong Wolf–Rayet star ($v_w = 2000 \mathrm{\,km\,s^{-1}}$, $\dot{M} = 10^{-4} \mathrm{\,M_\odot \, yr^{-1}}$) at a distance of $1\,\mathrm{pc}$ as a red, dashed line.
For the black line in figure \ref{fig:WindStrengthForPress} ($P_{ext} = 10^4 k_B\,\mathrm{K}\,\mathrm{ cm^{-3}}$) the difference to the O star is small but especially for a larger external pressure or smaller masses of the BES the Wolf-Rayet star is able to influence the stability of the BES.
This is important in more extreme environments like stellar clusters ($P_{ext} \approx 10^6 k_B\,\mathrm{K}\,\mathrm{ cm^{-3}}$ for evolved HII regions) in which also distances significantly smaller than $1\,\mathrm{pc}$ are common \citep{olivier2021evolution}.
Even larger external pressures can be found in young HII regions ($10^8 k_B\,\mathrm{K}\,\mathrm{ cm^{-3}} < P_{ext} < 10^{10} k_B\,\mathrm{K}\,\mathrm{ cm^{-3}}$, \cite{olivier2021evolution}) or in the galactic center ($P_{ext} > 10^9 k_B\,\mathrm{K}\,\mathrm{ cm^{-3}}$, \cite{burkert2012physics}).
Those regions are also associated with strong radiation fields and interacting stellar outflows of neighbouring stars which makes those systems more complex. 
Especially the interplay of evaporation and compression by those effects needs to be understood in greater detail before applying our criterion.
Another wind source are AGB stars with typical velocities of around $v_w \approx 10 \mathrm{\,km\,s^{-1}}$ and mass-loss rates of $\dot{M} = 10^{-7} \mathrm{\,M_\odot \, yr^{-1}}$ to $\dot{M} = 10^{-5} \mathrm{\,M_\odot \, yr^{-1}}$ \citep{hofner2018mass}.
Due to their lower wind velocities in comparison to O stars, they only have an influence in low-pressure regions ($P_{ext} \approx 10^4 k_B\,\mathrm{K}\,\mathrm{ cm^{-3}}$) and if their distance to the BES is significantly smaller than $1\,\mathrm{pc}$.\\
In the derivation of the criterion (\ref{eq:StabilityCriteriaPressure}) we explicitly assume a constant compression over time, i.e. it cannot be directly applied to time-dependent phenomena like the collision with a supernova remnant or the influence of bipolar outflows from protostars. 
In a future study, we will analyze how the criterion (\ref{eq:StabilityCriteriaPressure}) can be modified to also cover those cases.

\section{Summary} 
In this paper, we have analyzed the interaction of a Bonnor-Ebert sphere in hydrostatic equilibrium with the stellar wind of a nearby star. We concentrated on the low Mach regime for the wind so that the wind can be approximated by an additional external force and does not have to be modeled separately. In section \ref{sec:analyticDerivcation} we first derived analytically a stability criterion for the BES (see inequality (\ref{eq:StabilityCriteriaPressure})) and predicted that it should be valid for $2.5 < \xi_{max} <4.2$. In section \ref{sec:simulations} we presented simulations with GADGET-3 and were able to verify the criterion and also its validity range.
We showed that the efficiency of the wind to trigger a collapse strongly decreases for $\xi_{max} < 2.5$ and for $\xi_{max} < 2$ we were not able to form a sink particle. 
In section \ref{sec:discussion} we discussed the implications of our results for protostellar cores in molecular clouds. 
We find that winds have an influence in low-pressure regions and on large cores. The smallest core that can be destabilized for a typical external pressure of $10^5 k_B\,\mathrm{K}\,\mathrm{ cm^{-3}}$ by a wind has a mass of around $0.5\,\mathrm{M_\odot}$, while otherwise cores below $1.5 \,\mathrm{M_\odot}$ would be stable.\\
For the future, we plan further studies with the moving mesh code Arepo \citep{springel2010pur,weinberger2020arepo} which will allow us to directly model the ejecta of the wind.
We also plan to continue the simulations presented in this paper until the BES is completely dispersed and to analyze the mass distribution of the formed sink particles.

\acknowledgments
The authors acknowledge support from the Deutsche Forschungsgemeinschaft (DFG, German Research Foundation) under Germanys Excellence Strategy - EXC-2094 - 390783311 from the DFG Cluster of Excellence ’ORIGINS’. 
We made use of the SPLASH software package \citep{price2007splash} to visualize our simulations.

\bibliography{main}{}
\bibliographystyle{aasjournal}

\listofchanges
\end{document}